\documentclass[useAMS,usenatbib,aas_macros]{mnras}
\usepackage{graphicx}
\usepackage{amssymb}

\usepackage{epsfig}
\usepackage{color}
\usepackage{amssymb}
\usepackage{ams math}
\usepackage[greek,english]{babel}
\usepackage{subfig}
\usepackage{relsize}

\def\tdhst{3D$-$\emph{HST}}
\def\hst{\emph{HST}}

\def\nii{\mbox{[N\,{\sc ii}]}}
\def\sii{\mbox{[S\,{\sc ii}]}}

\def\halpha{\mbox{H\,{\sc $\alpha$}}}
\def\hbeta{\mbox{H\,{\sc $\beta$}}}

\def\niiha{\nii/\halpha}
\def\siiha{\sii/\halpha}

\def\hii{\mbox{H\,{\sc ii}}}

\def\oiiia{\mbox{[O\,{\sc iii]\sc{$\lambda$5007}}}}
\def\oiiiall{\mbox{[O\,{\sc iii]\sc{$\lambda$5007}}}}
\def\oiii{\mbox{[O\,{\sc iii]}}}

\def\oiiihb{\oiii/\hbeta}

\def\oii{\mbox{[O\,{\sc ii]\sc{$\lambda$3727}}}}

\def\tdhst{3D-$HST$}

\title[\mbox{\rm{[O}\,{\sc iii]}}/\hbeta \ evolution between $0<z<4$]{Changing physical conditions in star-forming galaxies between redshifts $0<z<4$: \oiiihb \ evolution}

\author[F. Cullen]{\parbox\textwidth{F. Cullen$^{1}$\thanks{E-mail:fc@roe.ac.uk}, 
M. Cirasuolo${^{2,1}}$, 
L.J. Kewley${^3}$,
R.J. McLure${^{1}}$, 
J.S. Dunlop${^{1}}$\\ 
and R.A.A Bowler${^{1}}$ }\\\\
$^{1}$SUPA\thanks{Scottish Universities Physics Alliance}, Institute for Astronomy, University of Edinburgh, Royal Observatory, Edinburgh EH9 3HJ \\
$^{2}$European Southern Observatory, Karl-Schwarzschild-Strasse 2, D-85748
Garching bei Muenchen, Germany\\
$^{3}$ Australian National University, Research School for Astronomy $\&$ Astrophysics, Mount Stromlo Observatory, \\Cotter Road, Weston, ACT 2611, Australia}

\begin{document}

\date{Accepted 2016 May 13}

\pagerange{\pageref{firstpage}--\pageref{lastpage}} \pubyear{2016}

\maketitle	

\label{firstpage}

%%%%%%%%%%%%%%%%%% ABSTRACT %%%%%%%%%%%%%%%%%%%
\begin{abstract}
We investigate the redshift evolution of the \oiiihb \ nebular emission line ratio for a sample of galaxies spanning the redshift range $0<z<4$. We compare the observed evolution to a set of theoretical models which account for the independent evolution of chemical abundance, ionization parameter and interstellar-medium (ISM) pressure in star-forming galaxies with redshift. Accounting for selection effects in the combined datasets, we show that the evolution to higher \oiiihb \ ratios with redshift is a real physical effect which is best accounted for by a model in which the ionization parameter is elevated from the average values typical of local star-forming galaxies, with a possible simultaneous increase in the ISM pressure. We rule out the possibility that the observed \oiiihb \ evolution is purely due to metallicity evolution. We discuss the implications of these results for using local empirical metallicity calibrations to measure metallicities at high redshift, and briefly discuss possible theoretical implications of our results. 
\end{abstract} 


%%%%%%%%%%%%%%%%% INTRODUCTION %%%%%%%%%%%%%%%%%

\section{Introduction}

In the local Universe, thanks to the large statistical galaxy sample provided by the Sloan Digital Sky Survey (SDSS), key physical properties of the star-forming galaxy population have been investigated for $\sim 10^5$ galaxies.
For example, the position of the star-forming `abundance sequence' of local galaxies in the \oiiihb \ vs \niiha \ diagram \citep[][hereafter BPT diagram]{baldwin1981} has become well established \citep[e.g.][]{kauffmann2003a,kewley2006}.
This abundance sequence is a key diagnostic of the physical conditions in star-forming galaxies due to its dependence on the global gas-phase metallicity distribution, the stellar ionizing radiation field and various properties of the ISM within galaxies such as the geometrical distribution of gas and electron density \citep[see e.g.][]{kewley2013a}. 

Since the BPT diagram is constructed from the strongest rest-frame optical emission lines, most readily observable at high redshifts, it is possible to trace the BPT abundance sequence across cosmic time.
Therefore, with new ground-based multi-object spectrographs such as KMOS \citep{sharples2013} and MOSFIRE \citep{mclean2012}, the BPT diagram is beginning to be studied with large statistical samples out to $z\gtrsim2.2$ \citep[e.g.][]{steidel2014, shapley2014}.
Interestingly, it has become apparent that the BPT abundance sequence defined by local galaxies evolves out to high redshift, suggesting an evolution in the physical conditions of star-forming regions \citep[][]{kewley2013b, steidel2014, shapley2014}.
Specifically, from the growing sample of galaxies at $z>1$, an evolution away from the local abundance sequence toward higher values of the \oiiihb \ and/or \niiha \ ratio is observed \citep[see e.g.][for discussions]{kewley2013b, cullen2014, holden2014, steidel2014, shapley2014, kewley2015}.
\citet{steidel2014} have definitively confirmed this abundance sequence trend using MOSFIRE observations of 179 star-forming galaxies at $z\sim 2.3$, showing that these galaxies form a distinct, yet similarly tight, locus in the BPT plane compared to SDSS galaxies \citep[see also][]{shapley2014, masters2014}.

The cause of the abundance sequence evolution is still not fully resolved. 
\citet{kewley2013a} have investigated theoretically how this evolution may be related to a change in physical conditions in galaxies at high redshift.
They suggest more extreme ISM conditions in high redshift \hii \ regions will result in the abundance sequence at a given redshift being offset from the one defined locally. 
However, the precise nature of these `extreme conditions' is still a matter of debate. 
\citet{kewley2013a} show how the independent effects of a harder ionizing radiation field, an increase in the reservoir of ionizing photons (more stars forming within a given volume) and an increasing electron density (or equivalently ISM pressure) can shift the position of galaxies in the BPT diagram in a similar way to that observed.
\citet{kewley2013b} conclude that star-forming galaxies at high redshift are consistent with having some combination of a larger ionization parameter, larger electron density, and/or harder ionizing radiation field.
\citet{steidel2014}, using an independent model, reach a similar conclusion, and also suggest that variations in the N/O ratio at fixed O/H in high redshift galaxies may also be responsible in part for the observed shift in locus in the BPT diagram.
\citet{sanders2016} argue that high redshift galaxies are consistent with having the same ionization parameter as local galaxies at fixed metallicity and that the primary driver of the BPT diagram offset is an elevated N/O ratio at fixed O/H.
Most recently, \citet{kashino2016} have concluded that the line ratio evolution is caused by a combination of an increase in the electron density and ionization parameter, where the increase in ionization parameter out to $z=1.7$.

In this paper, motivated by the high redshift BPT studies, we investigate the physical conditions in high redshift galaxies by comparing the redshift evolution of the \oiiihb \ nebular emission line ratio to theoretical models of star-forming galaxies.
The advantage of focusing solely on the \oiiihb \ ratio is two-fold.
Firstly, the evolution of \oiiihb \ is independent of any N/O variations, and thus mitigates any problems related to the degeneracy between evolving ionization conditions and evolving N/O.
Secondly, it also allows us to extend the observed data out to $z \sim 3.5$, beyond where the \niiha \ ratio can be measured with current instrumentation.
Recently, \citet{kewley2015} have shown that in the redshift range $0.2<z<0.6$ the \oiiihb \ rises by 0.2 - 0.3 dex, and attribute this evolution to an increase in the ionization parameter over this redshift range due to an larger ratio of young to old stars in galaxies.
In this paper, we are able to extend the \oiiihb \ analysis over a larger redshift baseline.

Primarily, we utilize an independent reduction of \tdhst \ survey data \citep{brammer2012_3dhst, momcheva2015} to constrain the line ratio evolution in the redshift range $1.3 < z < 2.3$.
With the \tdhst \ data we are able to account for any selection effects resulting from redshift-dependent line luminosity thresholds, the importance of which has been recently highlighted in the literature. 
For example, \citet{juneau2014} find that, using a sample of SDSS galaxies, it is possible to re-create the abundance sequence offset at high redshift by applying increasing \oiii \ and \halpha \ luminosity cuts to the sample, and emphasize the importance of comparing emission-line-luminosity matched samples when comparing local and high redshift galaxies \citep[see also][]{cowie2016}.
We address the issue of completeness and selection biases in the \tdhst \ sample in Section \ref{sec_data}.
Once these luminosity completeness effects within \tdhst \ have been accounted for, we are able to supplement these data with $z\simeq3.5$ observations from the AMAZE/LSD surveys \citep{maiolino2008, mannucci2009, troncoso2014} and a sample of galaxies from \citet{holden2014}.

We then compare the evolution of the \oiiihb \ ratio between the SDSS, \tdhst, and $z\simeq3.5$ samples, across $\sim 10.7$ Gyr of cosmic time, to the theoretical models introduced in \citet{kewley2013a}.
We describe how these models are applied in the context of the present work in Section \ref{sec_theory}.
These models allow different scenarios for the physical evolution of the star-forming regions with redshift to be distinguished.
Specifically, we are able to separate the effects on the line ratio due to global evolution in metallicity, ionization parameter and ISM pressure.
The results are presented in Section \ref{sec_oiiihb_redshift}.
Finally, in Section \ref{sec_discuss} we discuss the practical implication of the results with respect to accurately measuring the gas-phase metallicity of 
galaxies at high redshift and explore some theoretical interpretations of our results.

Throughout, we refer to the \oiiia \ nebular emission line as \oiii \ unless explicitly stated otherwise.
We assume a cosmology with $\Omega_m$ = 0.3, $\Omega_{\Lambda}$ = 0.7 and $H_0$ = 70 km s$^{-1}$ Mpc$^{-1}$.

\section{Data}\label{sec_data}

The data analysed in this paper are drawn from a variety of separate star-forming galaxy catalogues across the redshift range $0<z<4$.
The objective behind this selection is to sample the \oiiihb \ vs. redshift relation across the widest redshift baseline possible. 
Below we describe the separate samples. 
All datasets are cut to only include galaxies with $\mathrm{log}(M/M_{*})>9.0$, for consistent comparison with the theoretical models described in Section \ref{sec_theory}.

\subsection{\tdhst}

The primary dataset presented in this paper, in the sense that it is the largest dataset used to constrain the evolution of the \oiiihb \ ratio at $z \gtrsim 1$, is a sample of galaxies drawn from an independent reduction of the \tdhst \ survey \citep{brammer2012_3dhst, momcheva2015}.
Below we give details of the \tdhst \ sample selection and spectroscopic analysis relevant to this study; however see \citet{cullen2014} for a more detailed description of the data reduction process.

\subsubsection{\tdhst sample selection}

The \tdhst \ galaxy sample is drawn from three of the five survey fields (GOODS-S, COSMOS and UDS) and for the analysis presented in this paper we only consider galaxies in the redshift range $\rm 1.3 \le z \le 2.3$, where both the \oiii \ and \hbeta \ lines are observable in the grism spectra. 
All galaxies are selected visually from the reduced 2D and 1D spectra in each field. 
In the visual inspection we check each spectrum for contamination by nearby sources, rejecting emission-line spectra where the emission lines are severely contaminated, or the contamination model leaves clearly erroneous residual features in the continuum.
As described in \citet{cullen2014} we use existing CANDELS catalogues for the GOODS-S and UDS fields as described in \citet{guo2013} and \citet{galametz2013} respectively.
For COSMOS we produce a custom catalogue, the details of which are described in \citet{bowler2012}.
Stellar masses for the \tdhst \ galaxies are measured from rest-frame UV to mid-IR photometry using the publicly available code \textsc{lephare} \citep{ilbert2006}.
We run \textsc{lepahre} with solar-metallicity \citet{bruzual2003} templates assuming exponentially declining $\tau$ model star-formation histories with $\tau$ = 0.3, 1, 2, 3, 5, 10 , 15, 30 Gyr and a Chabrier initial mass function (IMF). The \citet{calzetti2000} attenuation law is used to account for dust extinction with $E(B-V)$ values ranging from 0 to 0.6. The age of the model is allowed to vary between 0.05 Gyr and the age of the Universe at the spectroscopic redshift of the galaxy.
We remove AGN contaminants and exclude galaxies with a $<$ 3$\sigma$ detection in \oiii \ as described below.
This leaves a final \tdhst \ sample of N=211 galaxies down to a minimum \oiii \ line flux of 1.2 x 10$^{-17}$ $\rm{ergs/cm^{2}/s}$.

\subsubsection{Continuum and line flux fitting}

For each individual galaxy in the sample we first fit the continuum by masking the emission lines in the grism spectra, then, using a least-squares minimization technique, we find the best-fitting normalization between the grism continuum and the best-fitting BC03 model returned from SED-fitting. 
The normalized BC03 model is then taken as the continuum of the galaxy and subtracted from the grism spectrum. This method of continuum subtraction accounts for the \hbeta \ absorption in the stellar continuum through the BC03 models.
We then fit the \oiii \ and \hbeta \ line using a triple gaussian function. 
To estimate the errors on line measurements we perturb each spectrum, at each wavelength pixel, by drawing a random number from a Gaussian distribution with a standard deviation equal to the error value in that pixel.
The continuum and line fits are then performed on the perturbed spectrum.
This process is repeated 10$^3$ times and, for each line (\oiii \ and \hbeta), the final line flux and error is taken as mean and standard deviation of the resulting distributions.

\subsubsection{AGN removal}\label{sec_agn_removal}

The theoretical models for \oiiihb \ evolution explored in this paper assume the ionizing radiation responsible for the nebular emission has a stellar origin.
Therefore, to facilitate an accurate comparison between models and data, care must be taken to account for the effect of AGN contamination in our sample. 
Unfortunately the wavelength coverage, sensitivity, and resolution of the grism spectra do not allow us to detect simultaneously enough emission lines to use classical AGN diagnostic diagrams (e.g. the BPT diagram).
Therefore, the most reliable method for AGN removal applicable to our sample is X-ray detection. 
We use the 4Ms GOODS-S \citep{xue2011} catalogue and 1.8Ms COSMOS \citep{civano2012} catalogue to remove N=17 X-ray selected AGN from our sample.
We do not apply a common threshold to account for the different catalog depths, N=12/17 and N=5/17 AGN were identified in GOODS-S and COSMOS respectively.
We could not perform a similar selection in UDS due to the lack of a sufficiently deep X-ray catalogue.
We choose not to apply other optical emission line diagnostics, for example the MEx AGN diagnostic \citep{juneau2014}, as the premise of this study is based on investigating the idea that conditions in star-forming regions become more extreme at high redshift.
Therefore, by using traditional optical methods we risk removing genuine extreme star-forming galaxies from our sample.
Moreover, \citet{coil2015} find that the MEx diagram does not robustly identify AGN at $z\sim2$.
Moreover, we find that only N=16/313 galaxies would be classified as AGN via the  MEx classification, and have checked that removing these galaxies has no effect on the final results.

In summary, we have identified and removed N=17 X-ray selected AGN ($\sim 5\%$ of the original sample).
This is slightly lower than, previous studies at similar redshifts \citep[e.g.][]{stott2013, zahid2014a} which find $\sim 10\%$ AGN contamination.
Based on the number of AGN identified in the GOODS-S x-ray catalogue, and assuming a constant number of AGN per unit area of sky, we estimate that our sample may be contaminated by $\sim$ 5 and 10 unidentified AGN in COSMOS and UDS respectively.
These numbers are also be consistent with the $\sim 10\%$ contamination fraction found in other studies.
Therefore, we cannot rule out residual AGN contamination in our final sample  primarily due to the lack of a sufficiently deep X-ray catalogue UDS and to insufficient line diagnostic tools.
However, we are able to account for this residual AGN contamination in our stacked spectra via a bootstrapping technique described in Section \ref{sec_stacking}.
Moreover we have verified that removing the N=68/211 UDS galaxies from our sample does not change any of the results presented in this paper.

\subsubsection{Stacking procedure}\label{sec_stacking}

To reduce the uncertainty in the measured line ratios due to the lower signal-to-noise of the weaker \hbeta \ line we must stack the \tdhst \ spectra.
Throughout this paper we stack the galaxies in redshift space.
A redshift bin is defined, and within each bin the galaxy spectra were interpolated to a common wavelength grid. 
We then determine the inverse variance weighted mean (or median) flux within each wavelength bin across all the galaxies in that bin. 

We estimate errors on line flux measurements in the stacked spectra by bootstrapping the data. 
Any large scatter induced by AGN contamination will also be accounted for by bootstrapping. 
For each stack we randomly select N galaxies, with replacement, where N is the number of galaxies in the stack.
We then recombine the spectra and measure the line fluxes as described above.
This process is repeated $10^3$ times and the flux and error are taken as the mean standard deviation of the final distribution for each line.

\subsection{SDSS}

When discussing the evolution of physical conditions in \hii \ regions with redshift, we require a comparison sample with which to estimate values of a given parameter (e.g. metallicity, ionization parameter, ISM pressure) typical of galaxies in the local Universe.
For the local galaxy sample we use a Sloan Digital Sky Survey (SDSS) catalogue containing N=51,262 star-forming galaxies from \citet{zahid2013}; this sample of star-forming galaxies is selected in the redshift range 0.02 $<$ $z$ $<$ 0.12, with an aperture covering fraction $\gtrsim$ 20 $\%$ to minimize aperture effects and with AGN removed using the \citet{kewley2006} optical classification scheme. 
The emission-line fluxes in this catalogue are taken from the JHU/MPA catalogues\footnote{http://www.mpa-garching.mpg.de/SDSS/DR7/}, where the line fluxes have been corrected for underlying stellar absorption.
Stellar masses for the SDSS sample are calculated from the SDSS $ugriz$ photometry as described in \citet{zahid2013}.

\subsection{AMAZE/LSD}

We utilize a sample of N=40 star-forming galaxies at ${3<z<4}$ (median $z=3.4$) taken from \citet{troncoso2014}. 
The data compiled in \citet{troncoso2014} were taken as part of the AMAZE \citep{maiolino2008} and LSD \citep{mannucci2009} surveys undertaken with the SINFONI instrument at the VLT. 
The galaxies were selected to exclude AGN based either on their UV spectra, X-ray data or MIPS $24\mu m$ flux (see \citet{maiolino2008} for details). 
We select galaxies from the \citet{troncoso2014} catalogue with a $3 \sigma$ detection in \oiii. 
The final sample contains N=33/40 galaxies.

\citet{troncoso2014} did not perform a correction for underlying \hbeta \ absorption, concluding the correction would be negligible due to the non-detection of a continuum in the majority of sources. 
Stellar masses for this sample were derived from either 14-band UV - \emph{Spitzer}-IRAC photometry or optical photometric data ($U,G,R,I$) plus \emph{Spitzer} IRAC and MIPS bands. 
This is the only sample with stellar masses derived using a Salpeter IMF \citep{salpeter1955} so we convert the quoted values to a Chabrier IMF \citep{chabrier2003} by dividing by 1.8.

\subsection{Holden et al. 2014}

We increase the $z>3$ sample size using the \citet{holden2014} dataset, consisting of N=18 Lyman-break selected star-forming galaxies at $z\sim3.2-3.7$ with \oiiihb \ ratios measured using the MOSFIRE instrument on the Keck-I telescope.
AGN contamination is ruled out based on the rest-frame UV spectra and X-ray data.
Again no correction has been made for the underlying \hbeta \ absorption, but again no spectra have reliably detected continua and thus, as argued by \citet{troncoso2014}, the correction would most likely be negligible.
We select galaxies from the \citet{holden2014} catalogue with a $3 \sigma$ detection in \oiii. 
The final sample contains N=15/18 galaxies. 
In combination with the AMAZE/LSD galaxies this gives a total of N=48 galaxies at $z>3$ with a median signal-to-noise ratio of 4 in the \oiiihb \ ratio.

\section{\oiii \ luminosity completeness}\label{sec_completeness}

Before investigating the \oiiihb \ versus redshift relation, it is important to understand any selection biases present in the above dataset which can affect the measurement of line ratios.
As has been recently pointed out by \citet{juneau2014}, any observed evolution of line ratios could be a result of larger line luminosity limits at high redshift.
Cosmological flux dimming will cause the minimum observable \oiii \ luminosity to increase with redshift, and if \oiiihb \ correlates with \oiii \ luminosity, an evolution to a higher ratio with redshift could be an artifact of the increasing \oiii \ luminosity detection threshold.

\subsection{[OIII] luminosity: evolution compared to SDSS}\label{sec_luminosity_ev_wrt_SDSS}

	\begin{figure}
		\centerline{\includegraphics[width=\columnwidth]{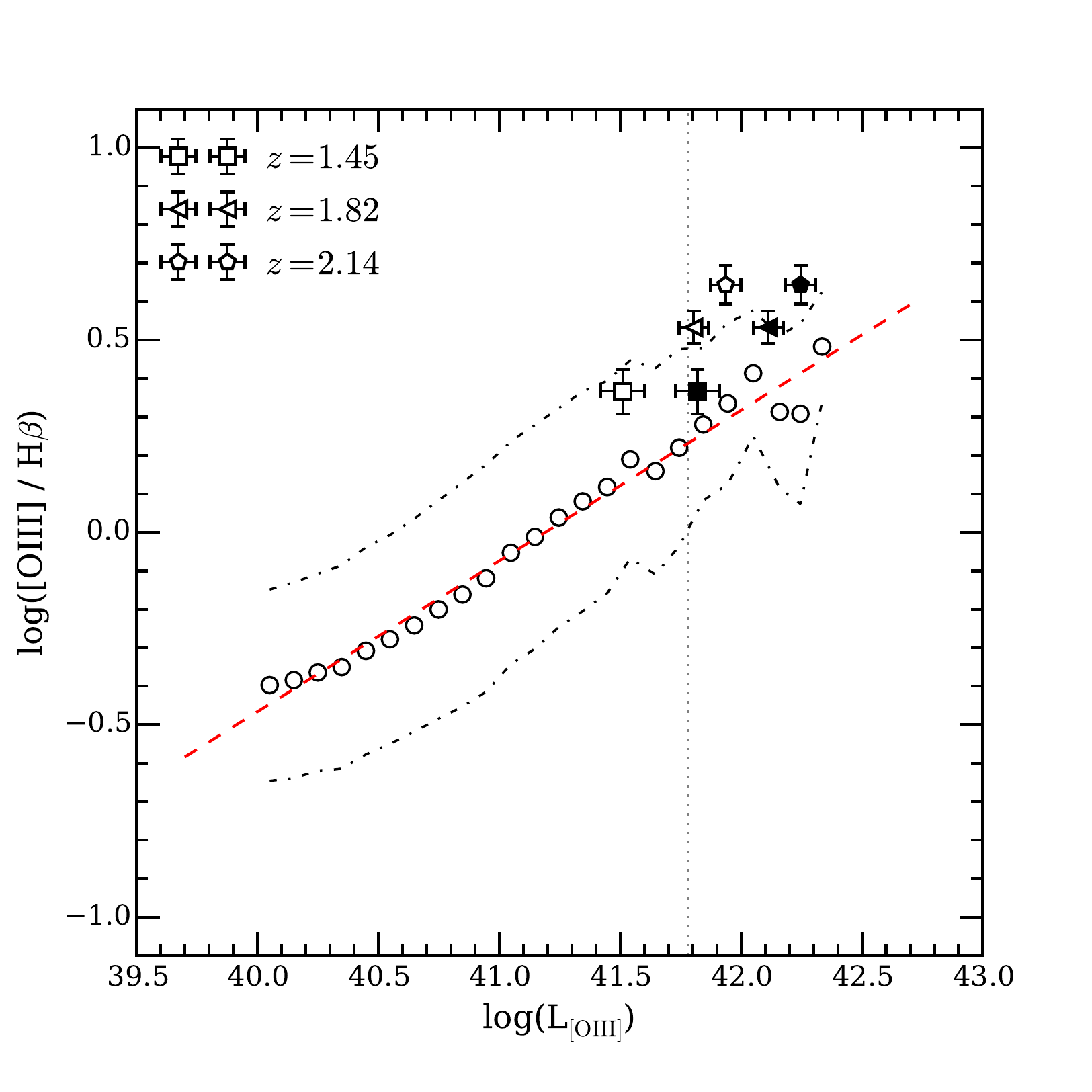}}
		\caption[\oiiihb \ ratio vs. \oiii \ luminosity for \tdhst \ and SDSS galaxies]{Relationship between \oiiihb \ and L$_{\rm{[OIII]}}$ for the SDSS sample compared to three \tdhst \ galaxy stacks.
		The small open circles show the averages of SDSS galaxies binned in 0.05 dex wide bins of log(L$_{\rm{[OIII]}}$), with the dot-dashed lines showing the $\pm 1 \sigma$ scatter.
		The red dashed lines shows a simple linear fit to the SDSS galaxies.
		The square, triangle and circle symbols show the position of the median stacked spectra from the \tdhst \ sample in increasing bins of redshift. 
		The open symbols show the positions of the stacks without applying a dust correction to L$_{\rm{[OIII]}}$, while the filled symbols show the same stacks after applying a \citet{calzetti2000} dust correction using the median $E(B-V)$ of the individual galaxies within the stack.
		The dotted vertical black lines shows the log(L$_{\rm{[OIII]}}$) completeness limit at $z>2$ for the \tdhst \ sample as described in Section \ref{sec_completeness}.} 
		\label{fig_sdss_oiii_lum}
	\end{figure}

We first explore how this affects any observed evolution compared to galaxies in the local Universe.
\citet{juneau2014} find, using an SDSS sample of $300,000$ star-forming galaxies, that by imposing a minimum line luminosity to \oiii \ and \halpha \, they can produce sub-samples in the SDSS data that mimic the BPT evolution observed at $z>1$. 
For the purposes of our results the implication is that as the line detection luminosity threshold is increased, an artificial evolution towards higher \oiiihb \ is observed.
We investigate whether the luminosity of \oiii \ correlates with \oiiihb \ in our SDSS sample in Fig. \ref{fig_sdss_oiii_lum}.
We find, in agreement with \citet{juneau2014}, that the \oiiihb \ increases with \oiii \ luminosity. 

Also plotted in Fig. \ref{fig_sdss_oiii_lum} are the position of three redshift stacks from the \ \tdhst \ sample both corrected (filled symbols), and uncorrected (open symbols), for dust using the \citet{calzetti2000} attenuation law.
The dust attenuation for the stack is taken as the median of the $E(B-V)$ values for each individual galaxy in the stack derived from the SED fitting.
We assume an extra extinction for nebular lines following the \citet{wuyts2013} prescription.
It can be seen from Fig. \ref{fig_sdss_oiii_lum} that, even in the case of no dust correction, our sample lies at the extreme end of the SDSS sample in terms of L$_{\rm{[OIII]}}$ and \oiiihb.
Fig. \ref{fig_sdss_oiii_lum} indicates that a comparison of the \tdhst \ sample to the total SDSS star-forming sample will be biased by the line luminosity detection limit of \tdhst.
Any observed \oiiihb \ evolution in this case could simply be an artifact of the clear correlation between L$_{\rm{[OIII]}}$ and \oiiihb.
Instead, we must take care to select galaxies from SDSS with comparable L$_{\rm{[OIII]}}$.
To establish the detection limit in \oiii \ luminosity in the \tdhst \ sample we perform a set of simulations as described below. 

\subsection{\tdhst \ simulations}\label{sec_completeness_sim_3dhst}

We use the highest-redshift galaxies in our \tdhst \ sample ($z>2$) to determine the \oiii \ luminosity threshold of the full sample. 
By using only galaxies with L$_{\rm{[OIII]}}$ above this level we can minimize any completeness biases which are introduced when comparing data at different redshifts.
To determine this minimum \oiii \ luminosity we run a set of simulations using the software package \texttt{aXeSIM}, a dedicated simulation package for \hst \ grism data.
We first create synthetic emission line spectra at $z=2.15$ (median redshift of the $z>2$ galaxies) spanning \oiii \ luminosities in the range $40.0<log(L_{\oiii})<43.0$ with \oiiihb $=5$, which is approximately the average ratio for the stacked galaxy spectra at $z>2$.
These synthetic spectra are then injected into a simulated G141 grism image with noise properties governed by the input exposure time to the simulation.

	\begin{figure}
		\centerline{\includegraphics[width=\columnwidth]{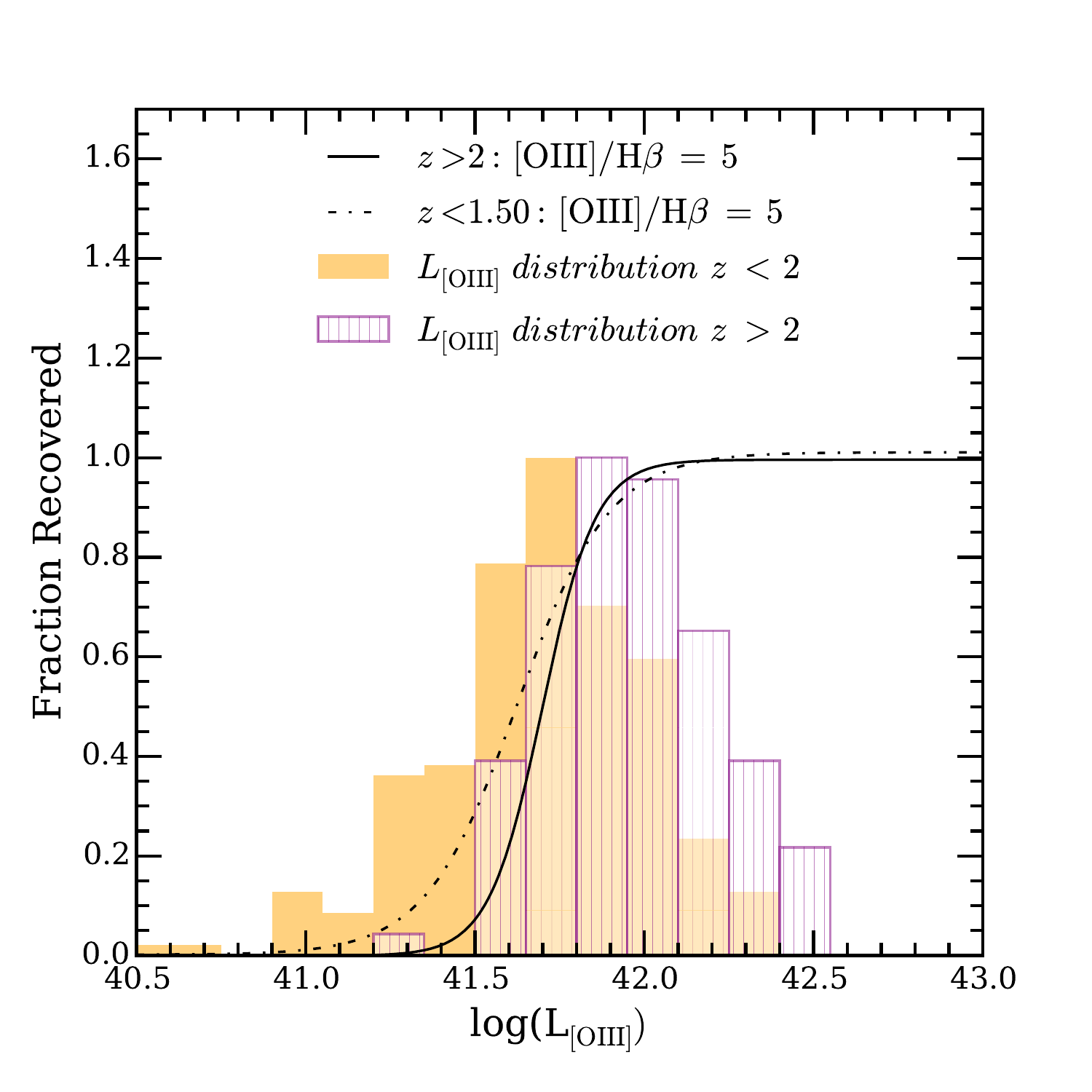}}
		\caption[\oiii \ luminosity completeness for the \tdhst \ sample]{Results of the \texttt{aXeSIM} simulations used to determine the completeness limits in log(L$_{\rm{[OIII]}}$) at $z>2$ for the \tdhst \ sample. The black solid line is the best fitting logistic function to simulation recovery fraction. The filled orange histogram shows the log(L$_{\rm{[OIII]}}$) distribution for the galaxies in the \tdhst \ sample at $z<2$ and the purple histogram shows the distribution for galaxies at $z\geq 2$. The black dot-dashed line shows the simulated recovery fraction of $z<1.5$ galaxies with high \oiiihb \ ratios (see text for a detailed explanation).}
		\label{fig_completeness}
	\end{figure}

Once the synthetic grism images are created, they are fed through the same \texttt{aXe} pipeline used for reducing the observed data, and 1D spectra extracted in the same way.
These spectra are visually inspected and either recovered or rejected in an identical way to the real data. 
From this we estimate the recovery fraction as a function of log(L$_{\rm{[OIII]}}$) and this is shown as the black solid line in Fig. \ref{fig_completeness}. 
From the fit to the recovery fraction shown in Fig. \ref{fig_completeness}, we infer 90$\%$, 75$\%$, 50$\%$ completeness limits at $z>2$ of, respectively, log(L$_{\rm{[OIII]}}$) $=$ 41.87, 41.77 and 41.70.
Also shown in Fig. \ref{fig_completeness} are the log(L$_{\rm{[OIII]}}$) distributions for the $z \geq 2$ (purple hatched histogram) and $z<2$ (orange filled histogram) galaxies in the sample normalized to a peak value of 1. 
It can be seen that, as expected, a significant fraction of galaxies at $z<2$ fall below the detection threshold at $z\geq 2$.
Therefore, for investigating the evolution of \oiiihb \ with redshift in  Section \ref{sec_oiiihb_redshift}, we limit all samples to include only galaxies at \oiii \ luminosities brighter than the above completeness limits.
Clearly, our results will be most robust for the 90$\%$ completeness cut, but it is also instructive to investigate how different completeness limits affect the measurements.

Another more subtle bias that could be introduced into our data is the non-selection of galaxies with high \oiiihb \ ratios at the lowest redshifts in the \tdhst \ sample.
Such a bias would artificially decrease the average \oiiihb \ at lower redshifts and could arise because the sensitivity of the G141 grism falls off at shorter wavelengths.
For any given redshift, at a large enough \oiiihb \ ratio, the \hbeta \ line will not be visible in the spectrum below a given \oiii \ flux.
Due to the decreased sensitivity at short wavelengths this flux limit will be lower at lower redshift.  
Without a visible \hbeta \ it becomes less likely the galaxy will be visually selected based on the single line detection of \oiii.
Therefore, galaxies with large \oiiihb \ may not be selected.
We run a further set of simulations to check whether galaxies with \oiiihb \ $=5$,  typical of our $z>2$ galaxies, would be visually selected at $z<1.5$.
The recovery fraction as a function of log(L$_{\rm{[OIII]}}$) is shown by the black dot-dashed line in Fig. \ref{fig_completeness}.
We find that above the completeness limit for $z>2$ galaxies discussed above, we are also complete in galaxies with high \oiiihb \ ratios at the lowest redshifts in our sample.

Finally we note that though our method of matching samples by \oiii \ luminosity provides the simplest way of ensuring no artificial evolution of the \oiiihb \ ratio, it would also be desirable to compare the evolution of the ratio in bins of various different physical properties.
However, we have confirmed that these luminosity cuts do not produce biases in the global properties of the SDSS sample with respect to the $z>2$ galaxies.
Specifically, the median masses of the SDSS and \tdhst \ samples (using the 90$\%$ threshold) are consistent (log(M/M$_\odot$) $=$ 9.36 and 9.49  respectively), as are the median SFRs (log(M$_\odot$yr$^{-1}$) $=$ 0.92 and 0.83 respectively).
The SDSS SFRs are derived from the \halpha \ luminosity, corrected for dust redenning using the \halpha/\hbeta \ ratio, and the \tdhst \ SFRs are taken from the best-fitting SED model.
Therefore, by applying an increasing \oiii \ luminosity cut to the SDSS sample we are selecting a population of low mass, highly star-forming galaxies similar to those observed at $z>1.5$.
Nevertheless, one way of extending this analysis in the future would be to select samples at different redshift matched in bins of various physical properties such as stellar mass, SFR, sSFR etc.
We intend to explore this topic in more detail in a future work.

To summarize, we run simulations to account for two sources of bias in the \tdhst \ sample: (i) an increasing lower limit to L$_{\rm{[OIII]}}$ as a function of redshift; (ii) possible non-detection of high \oiiihb \ ratio galaxies at the lowest redshifts.
To mitigate (i) we calculate the L$_{\rm{[OIII]}}$ limit at $z>2$ and apply this cut-off to all data which investigate the \oiiihb \ evolution in Section \ref{sec_oiiihb_redshift}.
We also confirm that, above the L$_{\rm{[OIII]}}$ limit imposed by this cut, the \tdhst \ sample is still complete for large \oiiihb \ ratios at the lowest redshifts.

\section{Theoretical Models}\label{sec_theory}

In Section \ref{sec_oiiihb_redshift} we compare the observed redshift evolution of the \oiiihb \ ratio to models described in a companion paper  (Kewley et al. in prep). 
The redshift evolution of the \oiiihb \ ratio is derived by combining stellar spectral synthesis and photoionization modeling of star-forming regions with a prescription for the chemical abundance evolution of galaxies with redshift.
We briefly describe these models here but refer readers to \citet{kewley2013a, kewley2013b} and Kewley et al. 2015 (in prep) for a detailed description.

To account for  the evolution of the gas-phase chemical abundance a prescription is taken from the cosmological hydrodynamic simulations of \citet{dave2011a, dave2011b}. 
A 3rd order polynomial is fitted to the change in chemical abundance ($\Delta$log(O/H)) between $z$ = $3$ and $z$ = $0$ for galaxies with $\rm{M}_{*}~>~10^{9}$ assuming momentum-driven winds, and is given by:

\begin{equation}
\label{eq_metallicy_ev_dave}
\begin{array}{rl}
\Delta\mathrm{log(O/H)}=-0.0013-0.2287z+0.0627z^2 \\
-0.0070z^3
\end{array}
\end{equation}

	\begin{figure}
		\centerline{\includegraphics[width=\columnwidth]{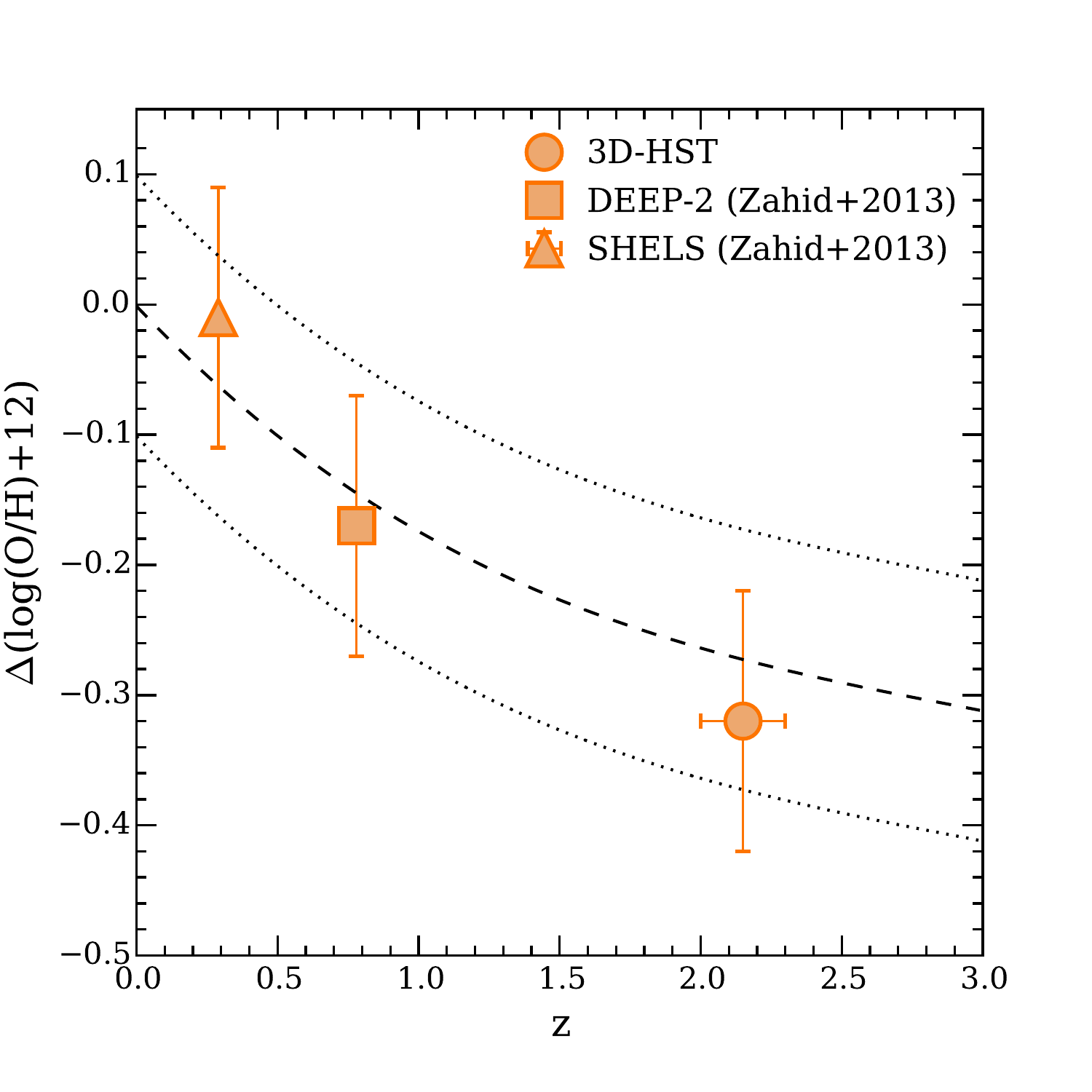}}
		\caption[Cosmic metallicity evolution between $z=3$ and $z=0$ from the \citet{dave2011a} models]{The assumed redshift evolution of the oxygen abundance used in the \oiiihb \ evolution models as described in the text. This evolution is taken from the cosmological hydrodynamic simulations of \citet{dave2011a, dave2011b} and is shown as the black dashed line, with the dotted line representing the $\sim$ 0.1 dex error. The three data points are from various observations where metallicities have been measured using the \citet{kobulnicky2004} calibration and averaged over all stellar  masses (triangle -  SHELS data from \citet{zahid2013}; square - DEEP2 data from \citet{zahid2013}; circle - \tdhst \ data from \citet{cullen2014}).}
		\label{fig_dave2011_metal_ev}
	\end{figure}

This equation clearly assumes a mass-invariant metallicity change for galaxies with $\rm{M}_{*}~>~10^{9}~\rm{M}_{\odot}$, and this may be a simplistic assumption.
However in Fig. \ref{fig_dave2011_metal_ev} we show that this evolution is a reasonable parameterization of the global metallicity evolution from z = 0 to z = 2.
In this figure we compare the the evolution in metallicity using data from \citet{zahid2013} and \citet{cullen2014} where, since Eq. \ref{eq_metallicy_ev_dave} is independent of stellar mass, the metallicity at a given redshift is calculated as an average over all stellar masses.
We take the line measurements from \citet{cullen2014} and re-calculate the metallicity using the \citet{kobulnicky2004} calibration for consistency with the \citet{zahid2013} data.

An EUV stellar radiation field is taken from the \citet{leitherer1999} Starburst99 stellar evolutionary synthesis models. 
The synthetic stellar spectra are modeled with the Pauldrach/Hillier model atmospheres \citep{hillier1998, pauldrach2001} including the effects of metal opacities. 
The resulting stellar radiation field is taken as the input ionizing source to the MAPPINGS IV photoionization code \citep{binette1985, sutherland1993, dopita2013} calculated over a large range in parameter space: metallicity, log(O/H) + 12 = [8.4, 8.7, 9.0, 9.2]; ionization parameter log($U$) = [$-$3.5, $-$3.0, $-$2.5, $-$2.0]; ISM pressure, log($P$/$k$) = [5.0, 5.5, 6.0, 6.5, 7.0].

To explore the evolution of the \oiiihb \ line ratios with redshift, four different scenarios are envisaged: (1) no evolution in the ISM pressure or ionization parameter (i.e. purely chemical abundance evolution), (2) only the ISM pressure evolves with redshift, (3) only the ionization parameter evolves with redshift, (4) both the ISM pressure and ionization parameter evolve with redshift.
Specifically, for the ISM pressure we assume a simple but extreme model in which ISM pressure increases by two orders-of-magnitude from $z$ $=$ $0.07$ (mean redshift of the SDSS) to $z$ $=$ $4$.
In an ionized gas with an isobaric density distribution the ISM pressure is simply determined by the electron density \citep{dopita2006} and recent work has shown that by $z\sim2.3$ the typical electron density in \hii \ regions has increased by one order of magnitude \citep{shirazi2014b,steidel2014, sanders2016}.
Crucially, it is unlikely that our model underestimates the effect of ISM pressure evolution.
The evolution of ionization parameter is modeled by an order-of-magnitude increase in the same redshift range.
This is consistent with recent measurements of ionization parameter out to $z\sim3$ \citep{nakajima2014}.
The equations for these models are given in Kewley et al. (in prep) along with further discussion of their motivation.
In each case, the reference values of ionization parameter and ISM pressure for SDSS galaxies is derived using on those galaxies which conform to the chosen \oiii \ luminosity threshold.

\subsection{Ionization parameter definition}

Conventionally, the ionization parameter is defined as the ratio of the number of ionizing photons passing through a unit area of ISM per second divided by the hydrogen atom density. 
The hydrogen atom density, in a fully ionized plasma, is approximately equivalent to the electron density.
However, since we are separately constraining the ionization parameter and ISM pressure (or electron density) in our models, it is important to note that the increasing ionization parameter models represent an increase \emph{at fixed ISM pressure}.
There are two ways to increase the ionization parameter in this way.
Firstly, the relative fraction ionizing photons can be increased by assuming a hotter stellar population which results in a harder EUV radiation field.
Secondly, the EUV radiation field can be scaled by, for example, having a larger number of stars forming within a \hii \ region.
Detailed explanations of these two effects are discussed in \citet{kewley2013a,kewley2015}.
In our models we increase the ionization parameter by scaling the radiation field and therefore the physical interpretation of a larger ionization parameter in this work is an increase in the number of hydrogen ionizing photons incident on a unit area of the ISM per second due to more stars forming within a given ISM volume.

Moreover it is worth noting that since the prescription for metallicity evolution given in Eq. \ref{eq_metallicy_ev_dave} is included in all models, the increase in ionization parameter due to a decrease in chemical abundance is naturally accounted for through the choice of lower metallicity Starburst99 stellar evolution models.
This anti-correlation between ionization parameter and chemical abundance has been observed locally \citep{perez_montero2014, sanchez2015}, and recently \citet{sanders2016} have evidence for a similar correlation at $z\sim2$.
Thus models 3 and 4 represent an increase in the ionization parameter over and above that expected due to the decrease in chemical abundance predicted from Equation \ref{eq_metallicy_ev_dave}.

\section{Redshift evolution of \oiiihb}\label{sec_oiiihb_redshift}

	\begin{figure*}
		\centerline{\includegraphics[width=4.8in]{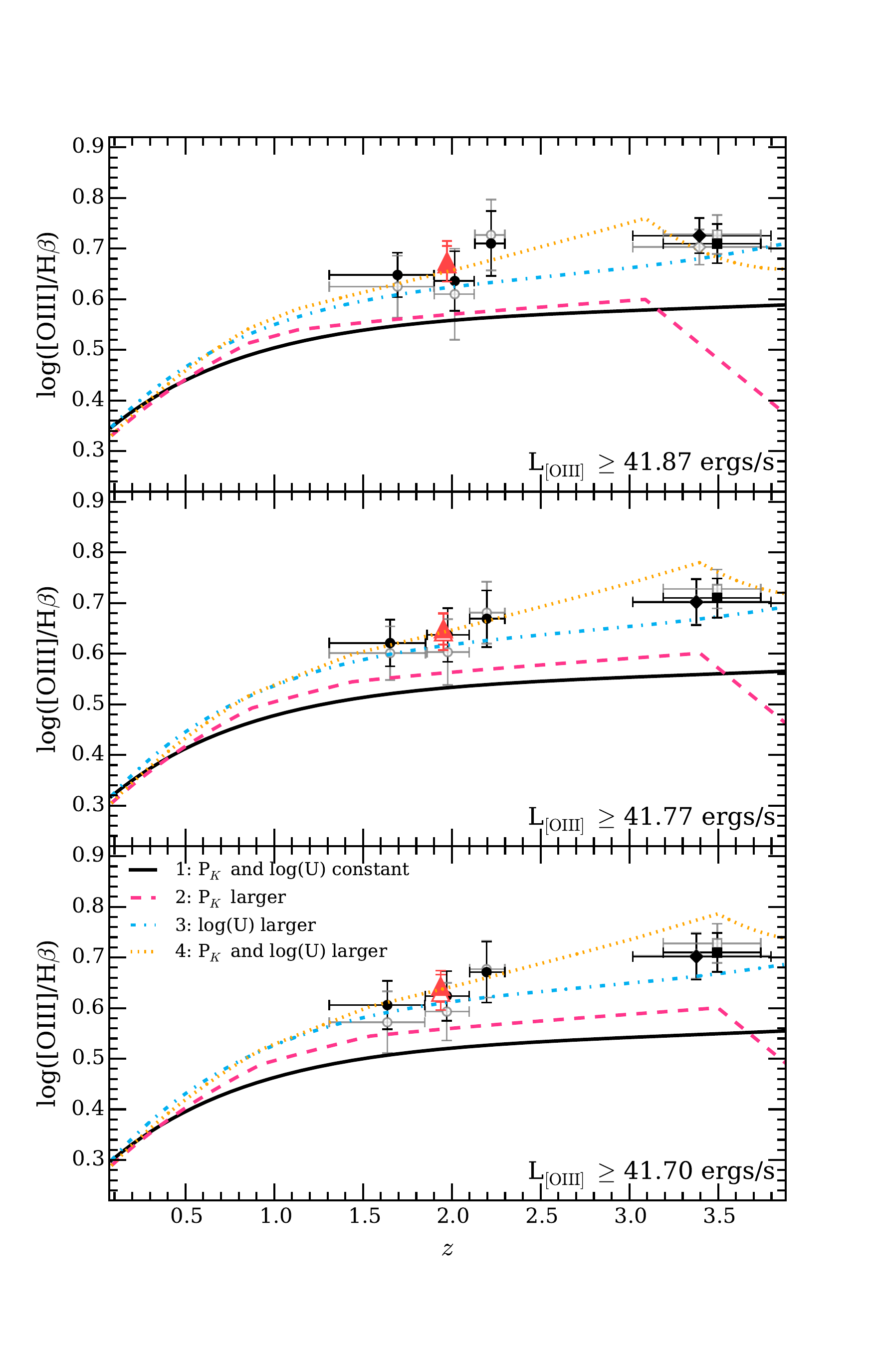}}
		\caption[\oiiihb / vs. redshift compared to the four theoretical evolution scenarios]{Comparison between the observed evolution of the \oiiihb \ ratio as a function of redshift and the theoretical models described briefly in Section \ref{sec_theory}. The circles show the \tdhst \ sample split over three redshift bins. the red triangles show the stack of all spectra in the \tdhst \ sample. The diamonds show the AMAZE/LDS data and the squares the \citet{holden2014} sample. The open and filled symbols correspond to an average and median stack, respectively (see text for details). Each panel shows a different \oiii \ luminosity cut as described in Section \ref{sec_completeness}.}
		\label{fig_oiiihb_redshift}
	\end{figure*}

We now compare the observed redshift evolution of the \oiiihb \ ratio to the theoretical models described above in each of the four scenarios. 
The models are run for \oiii \ luminosity limits of log$(L_{[OIII]})\geq$ 41.87, 41.77 and 41.70 \ as discussed above. 
In the four main datasets we utilize in the comparison (SDSS, \tdhst, AMAZE/LSD, Holden et al. 2014), we select the subset of each sample which adheres to the luminosity threshold.
Fig. \ref{fig_oiiihb_redshift} shows the \oiiihb \ vs. redshift relation for the four samples which, when combined, span the redshift range $0<z<4$. 
The SDSS data points are not explicitly shown in Fig. \ref{fig_oiiihb_redshift} but can be inferred from the $z=0$ origin of all the models. 

Each panel in Fig. \ref{fig_oiiihb_redshift} shows the results for each of the different luminosity cuts in \oiii.
In each panel the four model scenarios are plotted where the initial ionization parameter and ISM pressure has been calculated for the subset of SDSS galaxies which adhere to the stated luminosity threshold. 
For the \tdhst \ data we measure the \oiiihb \ ratio from both a mean and median stack in three redshift bins, as well as the mean and median of a stack of all the \tdhst \ spectra between $1.3<z<2.3$.
For the AMAZE/LSD and \citet{holden2014} data we simply plot the mean and median values of the individual \oiiihb \ ratios at the median redshift of the given samples.
All means are shown as filled symbols and all medians are shown as open symbols.

The first result of note is that the data rule out the model in which ionization parameter and ISM pressure remain constant (solid black curve), with only the metallicity of \hii \ regions evolving with redshift.
Therefore, an important result of this work is that the evolution of the \oiiihb \ line ratio, for the luminosity thresholds applied, cannot be accounted for purely by the evolution of metallicity.
This implies that the observed line ratios at high redshift are affected by the evolution in other physical parameters in \hii \ regions.
Importantly, this observation is independent of any variations in the N/O ratio.
An evolving N/O ratio has been suggested by some authors in order to explain emission-line ratios at high redshift, particularly in reference to the BPT diagram \citep[e.g.][]{masters2014, shapley2014, steidel2014, sanders2016}.
This is not to say that there is no N/O evolution, but rather that any N/O evolution must occur simultaneously with an evolution of other physical parameters.

The data also strongly disfavour the model in which ISM pressure evolves and ionization parameter remains constant (pink dashed curve).
For the most robust sub-set (90$\%$ completeness) only one data point is within 1$\sigma$ of this model, and across all panels the data are consistently in better agreement with either of the two remaining models.
Overall, the P/k evolution model is not sufficient to explain the majority of measured line ratios and, despite some of the individual measurements being formally consistent within 1$\sigma$, all are consistently systemically higher than the P/k evolution model prediction. 
Moreover, this model is clearly ruled by the stack across the full redshift range of the \tdhst \ spectra.

The remaining two models - ionization parameter only evolution (blue dot-dashed curve) and ionization parameter + ISM pressure evolution (yellow dotted curve) - are both favoured and are difficult to distinguish with the quality of data currently available.
However this still permits us to conclude from the models that, at a minimum, the ionization parameter of \hii \ regions is evolving with redshift. 
To recall, the ionization parameter is defined in our models as the number of hydrogen ionizing photons incident on a unit area of the ISM per second. 
It is not at present possible to conclusively say whether this is accompanied by an increase in the ISM pressure.
Particularly, in the redshift range $1.5<z<2.5$ covered by the \tdhst \ data, the two models are indistinguishable given the size of the uncertainties.
The only notable exception is the \tdhst \ stack across the full redshift range for the log$(L_{[OIII]})\geq$ 41.87 threshold, nevertheless no strong conclusions can be made on the basis of this one measurement.
At $z>3$ the models are more separated, however, over all luminosity limits, the AMAZE/LSD and \citet{holden2014} data still do not allow us to robustly distinguish between either scenario.
In Section \ref{sec_discuss} we discuss ways in which this situation can be improved with future observations.

To summarize, the data rule out the scenario in which the evolution of the \oiiihb \ line ratio is driven purely by metallicity evolution. 
In addition they require an evolution in the ionization parameter to higher values with redshift.
This rise in ionization parameter may also be accompanied by a simultaneous increase of the ISM pressure, however the data do not allow us to confirm or rule out ISM pressure evolution at this stage.
Possible theoretical explanations of this result are discussed in Section \ref{sec_discuss}.

\subsection{Comparison with the literature}

In this section we compare our results to a selection of recent observational studies which have also addressed the issue of evolving physical conditions in star-forming galaxies at high redshift.

\citet{sanders2016} have investigated ionization parameter and electron density evolution of galaxies out to $z\sim2.3$ based on a sample of $\sim200$ galaxies from the MOSDEF survey \citep{kriek2015}. 
Via a comparison with local galaxies across a wide range of line ratio diagnostic diagrams involving the \oiii, \oii, \halpha, \hbeta, \nii \ and \sii \ emission lines, Sanders et al. conclude that the typical electron density and ionization parameter increases between $z\sim0$ and $z\sim2.3$. 
Sanders et al. argue that the evolution to larger ionization parameters is solely a result of the lower metallicities of the high redshift galaxies, and that the ionization parameter of galaxies at fixed metallicity does not evolve with redshift.
In this scenario, the evolution of galaxies in the BPT diagram is primarily attributed to an increase in the N/O ratio at fixed O/H \citep[see also][]{masters2014, shapley2014}.
In terms of our models, the Sanders et al. conclusions would correspond to the pink dashed curve in Fig. \ref{fig_oiiihb_redshift}.
This is clearly inconsistent with the results presented here. 
However, while Sanders et al. rule out a scenario in which the ionization parameter increases due to a harder ionization spectrum, they do not investigate case, presented here, in which the ionization parameter is increased by scaling the ionizing spectrum.

\citet{kashino2016} have analysed the excitation properties a sample of 701 \halpha-selected galaxies at $1.4 \leq z \leq 1.7$ taken from the FMOS-COSMOS survey \citep{silvermann2015}.
Using line ratio diagnostics based around the \oiii, \halpha, \hbeta, \nii \ and \sii \ emission lines, Kashino et al. have concluded that the evolution of emission lines properties is due to both higher electron densities and ionization parameters at high redshift.
Furthermore, based on the suppression of the \siiha \ ratio at fixed \oiiihb, compared to local Universe galaxies, Kashino et al. conclude that the increase in ionization parameter is caused by a scaling up of the radiation field, as opposed to an increase in the hardness, in agreement with the results presented here.
However, Kashino et al. do not measure the \oii \ emission lines and are therefore unable to comprehensively address the N/O at fixed O/H issue.

At lower redshifts, \citet{kewley2015} have measured the evolution in the \oiiihb \ between $0.2 < z < 0.6$ using a sample of galaxies from the SHELS survey \citep{geller2016}.
Kewley et al. conclude that the evolution of the \oiiihb \ ratio at these redshifts is dominated by an increase in the ionization parameter caused by a scaling of the radiation field, ruling out either an increase in the hardness of the radiation field or an increase in electron density at these redshifts, based on measurements of \siiha \ and the \sii \ doublet ratio respectively.
Kewley et al. also demonstrate that this evolution is independent of stellar mass within the range $9.2 < \rm{log(M/M}_{\odot}\rm{)} < 10.6$.
Again the Kewley et al. results are consistent with the results presented here.

\section{Discussion}\label{sec_discuss}

In this section we discuss some of the practical and theoretical implications of the results presented here, as well as how our findings compare with results published in the recent literature.
If the evolution in ionization parameter and/or ISM pressure in star-forming regions at high redshift is real, this will have important consequences both for how strong-line ratios are applied in high redshift galaxies to measure physical parameters and, perhaps more fundamentally, on our theoretical understanding of the nature of \hii \ regions and the stars that power them.

\subsection{Impact of ionization parameter on metallicities inferred from strong line ratios}

        \begin{figure*}
        \centerline{\includegraphics[width=6.5in]{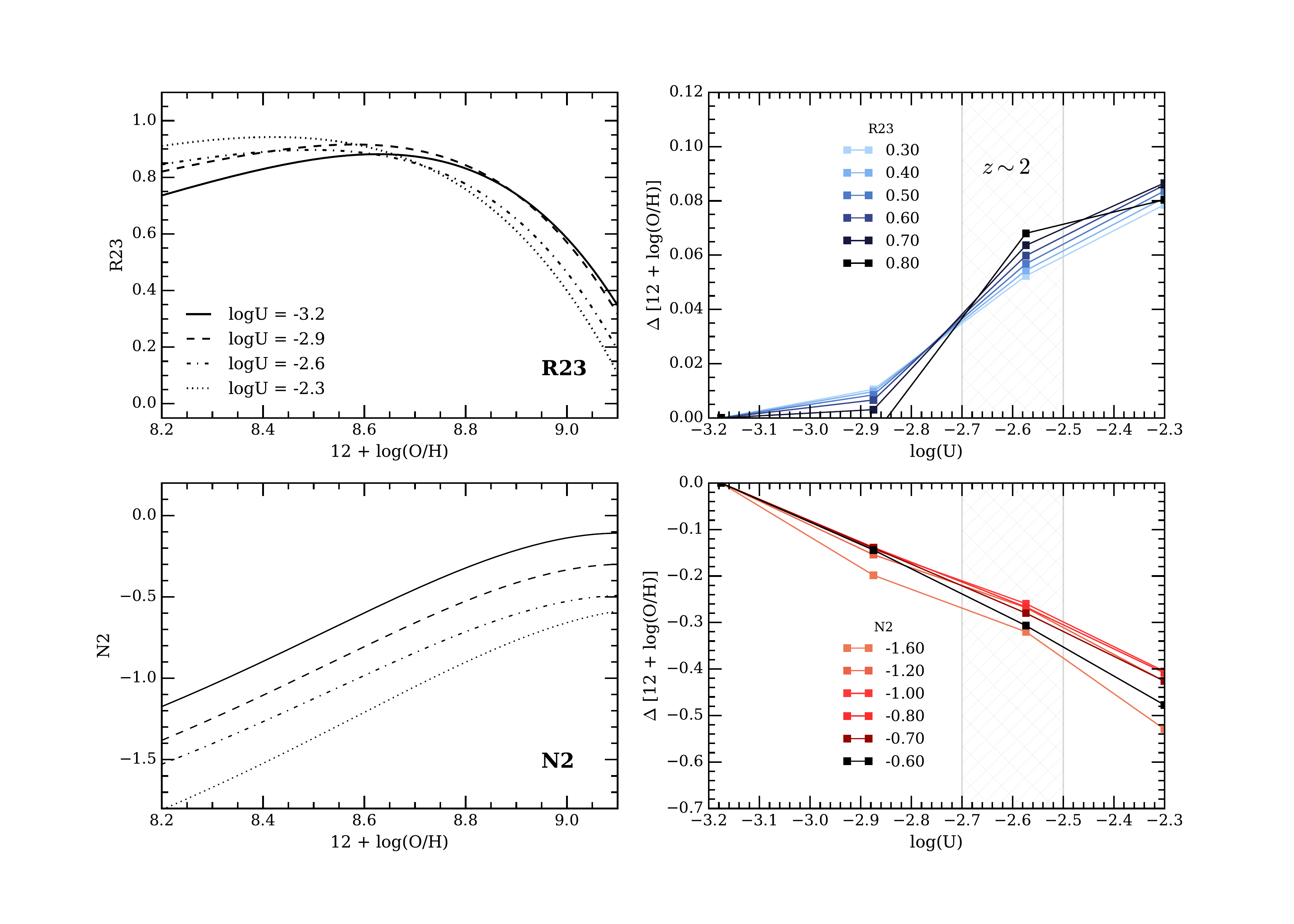}}
        \caption{The difference in inferred metallicity as a function of ionization parameter for different values of the R23 (top row) and N2 (bottom row) emission line ratios taken from \citet{kewley2002}. The differences are computed relative to the metallicity inferred assuming an ionization parameter typical of local star-forming galaxies. In each row the left-hand panel shows the line ratio metallicity calibration at four different values of the ionization parameter; the right-hand panel shows the metallicity differences as a function of ionization parameter for six different values of the ratios. The grey hatched regions represent values of ionization parameter typical of $z\gtrsim2$ galaxies.}
        \label{fig_metallicity_difference}
    \end{figure*}

In Fig. \ref{fig_metallicity_difference} we show an illustrative example of how failing to account for an evolving ionization parameter will bias metallicity diagnostics based on strong emission line ratios.
We use two commonly adopted emission lines diagnostics for the illustration: R23 $=$ (\oii + \oiiiall)/\hbeta \ (top row) and N2 $=$ \nii/\halpha \ (bottom row).
The left-hand panel in each row shows the metallicity calibration for four different values of the ionization parameter taken from \citet{kewley2002}.
It can be seen that, depending on the value of the ionization parameter adopted, the metallicity solution can be significantly different for both the R23 and N2 diagnostics.

This is explicitly shown in the right-hand panels of each row, where we show the difference in inferred metallicity ($\Delta$[12 + log(O/H)]) as a function of ionization parameter for different values of the two line ratios\footnote{For the R23 diagnostic we assume all galaxies fall on the upper branch (log(O/H)+12 $\gtrsim~8.6$) for simplicity.}.
The values of $\Delta$[12 + log(O/H)] are measured relative to a reference ionization parameter value of log(U)=$-3.2$ typical of SDSS galaxies \citep{liang2006, kewley2008}. 
Positive values indicate that the metallicity inferred from the line ratio will be overestimated if the typical SDSS ionization parameter is assumed and vice versa.
The ionization parameter values typical of current measurements of $z\sim2$ galaxies both from the \citet{cullen2014} \tdhst \ spectra at $z>2$, and the $z\sim2-3$ LBGs from \citet{nakajima2014} is shown as the grey hatched region.

Across the range of R23 and N2 ratios shown in the figure, the inferred metallicity for a given ratio becomes increasingly offset with increasing ionization parameter.
For the R23 ratio the difference can be of the order +0.06 dex at values of the ionization parameter typical of $z\sim2$ galaxies, while for N2 the difference is more pronounced, with offsets of order -0.3 dex.
Clearly the N2 ratio will significantly underestimate the metallicity of galaxies at $z\sim2$ if an ionization parameter typical of local galaxies is assumed.
Interestingly most observational studies have concluded that the N2 ratio overestimates metallicities by up to factor 3 at high redshift \citep[e.g.][]{liu2008,newman2014, steidel2014}.
This would suggest that other factors, such as increasing ISM pressure, the presence of shocks and/or an evolving N/O ratio may also be affecting this line ratio at high redshift.
On the other hand, the offset of the R23 indicator is comparable to the estimated uncertainties due to modeling inaccuracies \citep[$\sim$ 0.1 dex][]{kewley2002}, however the R23 indicator suffers from other significant systematic issues, namely the choice of upper or lower branch and sensitivity to the dust correction.
Fig. \ref{fig_metallicity_difference} illustrates how not accounting for the ionization parameter of a galaxy can have a large impact on the inferred metallicity, particularly for calibrations based on the \niiha \ ratio.
We therefore recommend that, in order to obtain reliable metallicities at high redshift, ionization parameter and metallicity should be simultaneously solved for.

\subsection{Improving the model constraints at high redshift}

\begin{figure}
		\centerline{\includegraphics[width=\columnwidth]{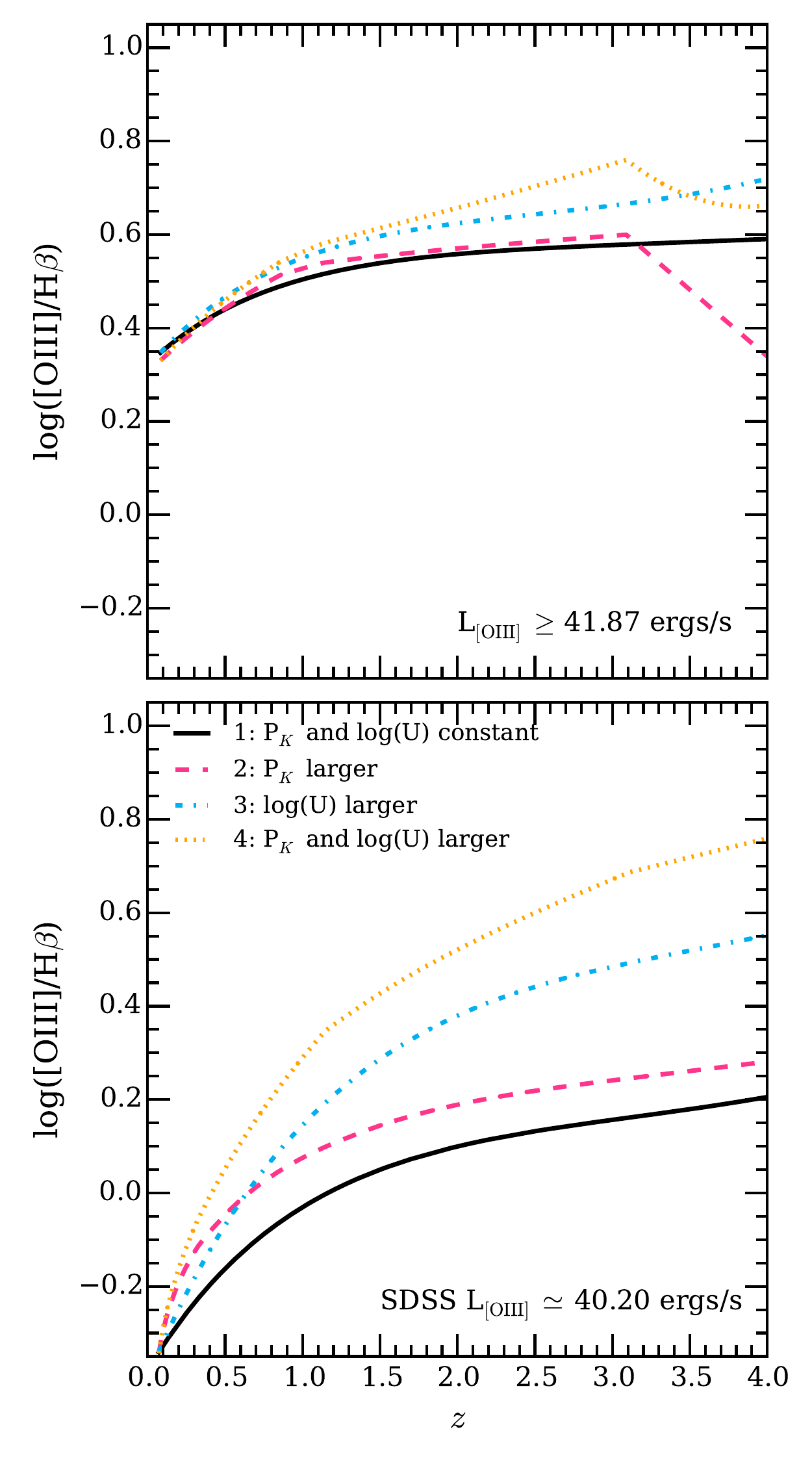}}
		\caption[Illustrating the $\mathrm{[OIII]}$ luminosity dependence of the theoretical models]{The impact of \oiii \ luminosity thresholds on the theoretical curves. The curves are plotted as in Fig. \ref{fig_oiiihb_redshift}. The top panel shows the models for the 90$\%$ completeness threshold used in this paper and the bottom panel shows the case for the lower luminosity limit of the full SDSS sample (log(L$_{\mathrm{[OIII]}}$) $\sim$ 40.2 ergs/s). It is clear from the figure how deeper observations will be better able to differentiate between models at all redshifts.}
		\label{fig_model_dependence_on_OIII_luminosity}
	\end{figure}

As discussed in Section \ref{sec_oiiihb_redshift}, with the data currently available at high redshift it is not possible to robustly distinguish between the models for ionization-parameter only evolution and ionization-parameter + ISM pressure evolution.
It is perhaps worth illustrating how this situation can be improved with future observations.
As previously discussed, the theoretical curves are dependent on the luminosity thresholds applied to the SDSS data, as this determines the reference values of ISM pressure and ionization parameter used as input to the models.
Fig. \ref{fig_model_dependence_on_OIII_luminosity} shows the model evolution from z = 0 to z = 4 for two cases; the top panel shows the \tdhst \ 90$\%$ completeness luminosity threshold and the bottom panel shows the model evolution when the full SDSS sample is used to calculate the initial values of the physical parameters (corresponding to log(L$_{\mathrm{[OIII]}}$) $\sim$ 40.2 ergs/s).

Fig. \ref{fig_model_dependence_on_OIII_luminosity} illustrates how raising the luminosity threshold makes it increasingly more difficult to distinguish between the models.
There is a much larger \oiiihb \ separation between the models when the lower luminosity threshold of full SDSS sample is used.
Therefore, deeper spectroscopic observation across a wider $\mathrm{L_{[OIII]}}$ baseline will allow a more robust distinction between the two currently indistinguishable scenarios.
Unfortunately SDSS-depth surveys at $z>1$ are not currently possible.
Nevertheless, state-of-the-art ground-based near-IR surveys such as MOSDEF will able to improve the situation by pushing down to 5$\sigma$ detections of log(L$_{\mathrm{[OIII]}}$) = 41.5, 41.0 ergs/ at $z=2.3,1.5$ respectively \citep{kriek2015}.
Even deeper spectroscopic observations will then be possible using future ground and space-based near-IR instruments such as $JWST$ and E-ELT.

\subsection{Theoretical Implications}\label{sec_theortical_imps}

Finally, we briefly discuss some of the possible theoretical explanations of the results presented here. 
Since our data favour models in which at least the ionization parameter increases with redshift, but cannot distinguish whether this is also accompanied by a simultaneous increase in ISM pressure, we limit our discussion to the possible causes of a higher ionization parameter.

An increased ionization parameter requires, from the models, an increased number of ionizing photons per unit area per unit time incident on the ISM.
One way of achieving this is to scale the ionizing radiation field by raising the star-formation rate within a given ISM volume. 
It is known that the average star-formation rate of galaxies increases out to $z\sim 2$ and beyond \citep[e.g.][]{whitaker2012, pannella2014}, if these stars form in similar volumes to local galaxies there will be an excess of ionizing photons at high redshift, and consequently a higher ionization parameter. 
Such dense star-forming clusters are commonly referred to as `super star clusters'. 
They have been theoretically investigated \citep[see e.g.][]{elmgreen1998, hopkins2013} and have been observed in local dwarf galaxies and galaxy mergers \citep[e.g][]{shioya2001,wofford2014}. 
However to date no comprehensive statistical study of the ionization parameter in these clusters has been made, nor any detailed investigation into their prevalence at high redshift.

The effective ionization parameter is also affected by the geometrical distribution of the gas in HII regions, for example clumpy HII regions allow more ionizing photons to escape and hence lower the effective ionization parameter \citep{nakajima2014}.
The difference between a radiation bounded HII region (where the nebula ends where hydrogen is completely recombined) and a density bounded HII region (where the entire nebula is ionized by the stars) is also important. 
In a density bounded HII region the \oiii \ zone (closest to the star) will likely remain unaffected but the Hydrogen recombination zone (which extends to larger radii) could be shortened, thus resulting in an increase in the \oiiihb \ ratio. Some evidence exists for density bounded HII regions at high-z \citep[see][for a detailed discussion of the geometrical effects]{kewley2013a}.

\section{Summary and Conclusions}

We have investigated the \oiiihb \ vs redshift relation for a sample of galaxies in the redshift range $0<z<4$ by combining data from SDSS, \tdhst, AMAZE/LSD and \citet{holden2014}.  
We have compared the \oiiihb \ evolution to the theoretical models described in \citet{kewley2013a}, that account for the effects of an evolving ionization parameter and ISM pressure, in order to further understand the evolution in the physical conditions of star-forming galaxies over the last $\sim11$ billion years of cosmic time.
We have explored how line luminosity selection limits can affect any observed trends and carefully accounted for these selection biases in all datasets.
We have finally discussed some of the practical and theoretical implications of the findings. 
Below is a summary of the results.

\begin{enumerate}

	\item We first show, using the SDSS sample, that there is a positive correlation between \oiiihb \ ratio and \oiii \ luminosity, as pointed out by \citet{juneau2014}. 
	This relationship could potentially bias any results since the observed evolution of the \oiiihb \ ratio could simply be an artifact of the increasing line luminosity threshold with redshift. 
	Therefore we conclude that the luminosity detection limit of all samples must be carefully accounted for when comparing to models of line-ratio evolution.

	\vspace{3mm}

	\item We correct for any biases introduced from an \oiii \ luminosity detection limit by restricting all samples to galaxies with log(L$_{\mathrm{[OIII]}}$) $>$ 41.87, 41.77 and 41.70, corresponding to completeness limits in the \tdhst \ sample of, respectively, 90$\%$, 75$\%$ and 50$\%$. 
	We show that, after applying these luminosity thresholds, the \tdhst \ stacks are consistent with an increase in the \oiiihb \ ratio at $1.3<z<2.3$ relative to SDSS galaxies. 
	Therefore we conclude this evolution is real and unaffected by the luminosity detection bias highlighted in \citet{juneau2014}.

	\vspace{3mm}

	\item We compare the \oiiihb \ redshift evolution to theoretical models described in Section \ref{sec_theory} based on the models of Kewley et. al. 2015 (in prep). 
	From these models we rule out the possibility that the line ratio increase is a direct result of metallicity evolution.

    \vspace{3mm}

	\item We find the observed line ratio evolution is best accounted for by, at least, an increase in the ionization parameter with redshift. In the context of our models this ionization parameter increase is a result of scaling up the incident EUV radiation field.
	Unfortunately the data cannot yet distinguish whether this ionization parameter increase is also accompanied by an increase in ISM pressure.
	These conclusions are independent of the line luminosity limits imposed.

	\vspace{3mm}

	\item Finally we discuss how, assuming an increase in ionization parameter and ISM pressure at high redshift, locally calibrated metallicity diagnostics will be biased when applied to high redshift galaxies.
	This highlights the need for the proper treatment of other physical parameters (and certainly the ionization parameter) when estimating the metallicity of high redshift galaxies.

\end{enumerate}

\section{Acknowledgments}
We would like to thank the referee for very helpful comments which have improved the manuscript.
FC and MC acknowledge the support of the Science and Technology Facilities Council (STFC) via the award of an STFC Studentship and an STFC Advanced Fellowship, respectively. 
RJM acknowledges the support of the European Research Council via the award of a Consolidator Grant (PI McLure). 
JSD acknowledges the support of the European Research Council via the award of an Advanced Grant, and the contribution of the EC FP7 SPACE project ASTRODEEP (Ref. No. 312725). 
This work is based on observations taken by the \tdhst \ Treasury Program (GO 12177 and 12328) with the NASA/ESA $HST$, which is operated by the Association of Universities for Research in Astronomy, Inc., under NASA contract NAS5-26555. 
This work is based (in part) on observations made with the Spitzer Space Telescope, which is operated by the Jet Propulsion Laboratory, California Institute of Technology under a contract with NASA. 
This research made use of Astropy, a community-developed core Python package for Astronomy \citep{astropy2013}, NumPy and SciPy \citep{oliphant2007}, Matplotlib \citep{hunter2007}, {IPython} \citep{perez2007} and NASA's Astrophysics Data System Bibliographic Services.

%%%%%%%%%%%%%%%%%%%% BIBLIOGRAPHY  %% %%%%%%%%%%%%

%\begin{thebibliography}{99}
\bibliographystyle{mnras}                      % The reference style
\bibliography{oiiihb}       % Multiple bib files.
%\include{../library.bib}
%\end{thebibliography}

\label{lastpage}

\bsp

\end{document}